\documentclass[11pt,twoside]{article}
\usepackage{asp2010}
\usepackage{natbib}
\bibpunct{(}{)}{;}{a}{}{,}
\resetcounters

\begin{document}

\title{MIDI interferometry of massive YSOs: Updates on the MPIA programme}
\author{Hendrik Linz$^1$, Roman Follert$^1$, Paul A.~Boley$^1$, Roy van Boekel$^1$, Bringfried Stecklum$^2$, Christoph Leinert$^1$, and Thomas Henning$^1$ 
\affil{$^1$Max--Planck--Institut f\"ur Astronomie MPIA, K\"onigstuhl 17, 
       D--69117 Heidelberg, Germany}
\affil{$^2$Th\"uringer Landessternwarte, Sternwarte 5, D-07778 Tautenburg,
       Germany}}

\begin{abstract}
Information about the inner structure of young stellar objects is crucial for understanding how the central forming stars gain their mass. However, especially for observations of (usually further away) high-mass young stellar objects, conventional imaging has limitations in spatial resolution. By means of mid-infrared interferometry, we can peek deeply into the strongly extinguished central 100 AU of such objects. Here, we report on new data we obtained within our programme using MIDI at the VLTI. Shown are preliminary results on the known outflow sources AFGL 2136 IRS 1 and Mon R2 IRS 3A. In particular, we describe how quantities like differential phases give additional geometrical structure information. We demonstrate how the combined interpretation of single--telescope and interferometric data at different wavelength regimes can lead to a more complete picture of the nature of such MYSOs.
\end{abstract}

\section{Introduction}
High-mass stars predominantly form in clustered environments much more distant than typical well-investigated low-mass star-forming regions. Thus, high spatial resolution is a prerequisite for making progress in the observational study of high-mass star formation. Furthermore, all pre-main sequence phases are usually deeply embedded. This often forces observers of embedded massive young stellar objects (MYSOs) to move to the mid-infrared (MIR) where the resolution of conventional imaging is limited to $>$0\farcs25 even with 8-m class telescopes. Hence, one traces linear scales still several hundred AU in size even for the nearest MYSOs, and conclusions on the geometry of the innermost circumstellar material remain ambiguous. A versatile method to overcome the diffraction limit of single telescopes is to employ interferometric techniques. Current MIR interferometry with most of the current facilities does not provide closure phases for direct image reconstruction. However, probing smaller spatial scales of $<0\farcs1$ by means of visibility and correlated flux measurements is a major step forward compared to the previous more or less unresolved thermal infrared imaging with single telescopes in the pre--ELT era for these sources. Already with this new structure information at hand, models of MYSOs can be advanced, and certain ambiguities, often arising from simple SED fitting of spatially unresolved photometry data, can be lifted. We also refer to the contribution by Hummel et al. in these proceedings for more details on infrared interferometry of MYSOs.

\section{The MPIA programme for mid--infrared interferometry of massive YSOs using MIDI}
In 2004, the MPIA has started a large programme to observe a variety of high--mass YSOs with ESOs mid--infrared interferometer MIDI \citep{2003SPIE.4838..893L} which had been designed and built at MPIA in cooperation with institutes in the Netherlands and France as well as ESO. In total, we have successfully observed 15 sources so far. At the beginning of the programme, all chosen targets were well--known infrared--bright sources that often are categorised as BN-type objects \citep[e.g.][]{1990FCPh...14..321H}, after the prototypical BN object in Orion. More recently, we have added both younger, more embedded sources \citep[HMPOs][]{2007prpl.conf..165B}, as well as more evolved sources and disk candidates, e.g., early (pre--)Herbig Be stars, to our list. All observed sources are clearly resolved with the interferometer baselines we applied ($\ge$16 m). A first status report has been given in \citet{2008JPhCS.131a2024L}, and individual objects have been treated in elaborate detail in \citet{2009A&A...505..655L} and \citet{2010A&A...522A..17F}.

\section{New results for selected objects}

We present here preliminary MIDI results for another two massive YSOs from our list. Both objects with luminosities equivalent to the early B-star regime, are well--studied previously, especially in the near-- and mid--infrared, and might drive molecular outflows. Therefore, deviations from spherical symmetry for the central sources especially on small spatial scales are expected. New MIDI results about a third object from our programme, AFGL 4176, which shows no clear indications for an outflow, but which attains a higher luminosity in the late O-star range, is the topic of the contribution by Boley et al.~in these proceedings.

\subsection{AFGL 2136 IRS 1}\label{Sect:AFGL2136}

\begin{figure}[t]
 \includegraphics[width=6.7cm]{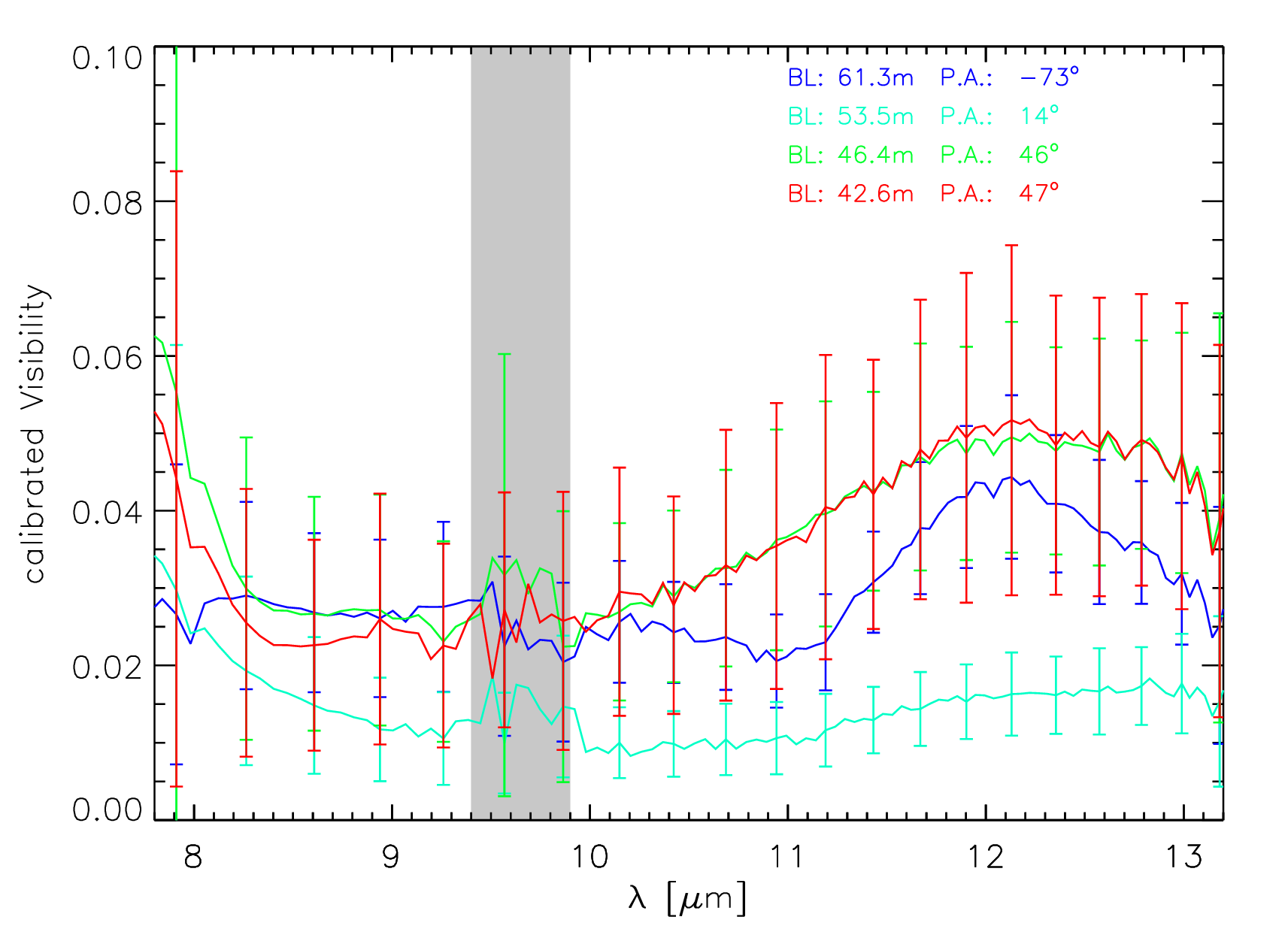}\includegraphics[width=6.7cm]{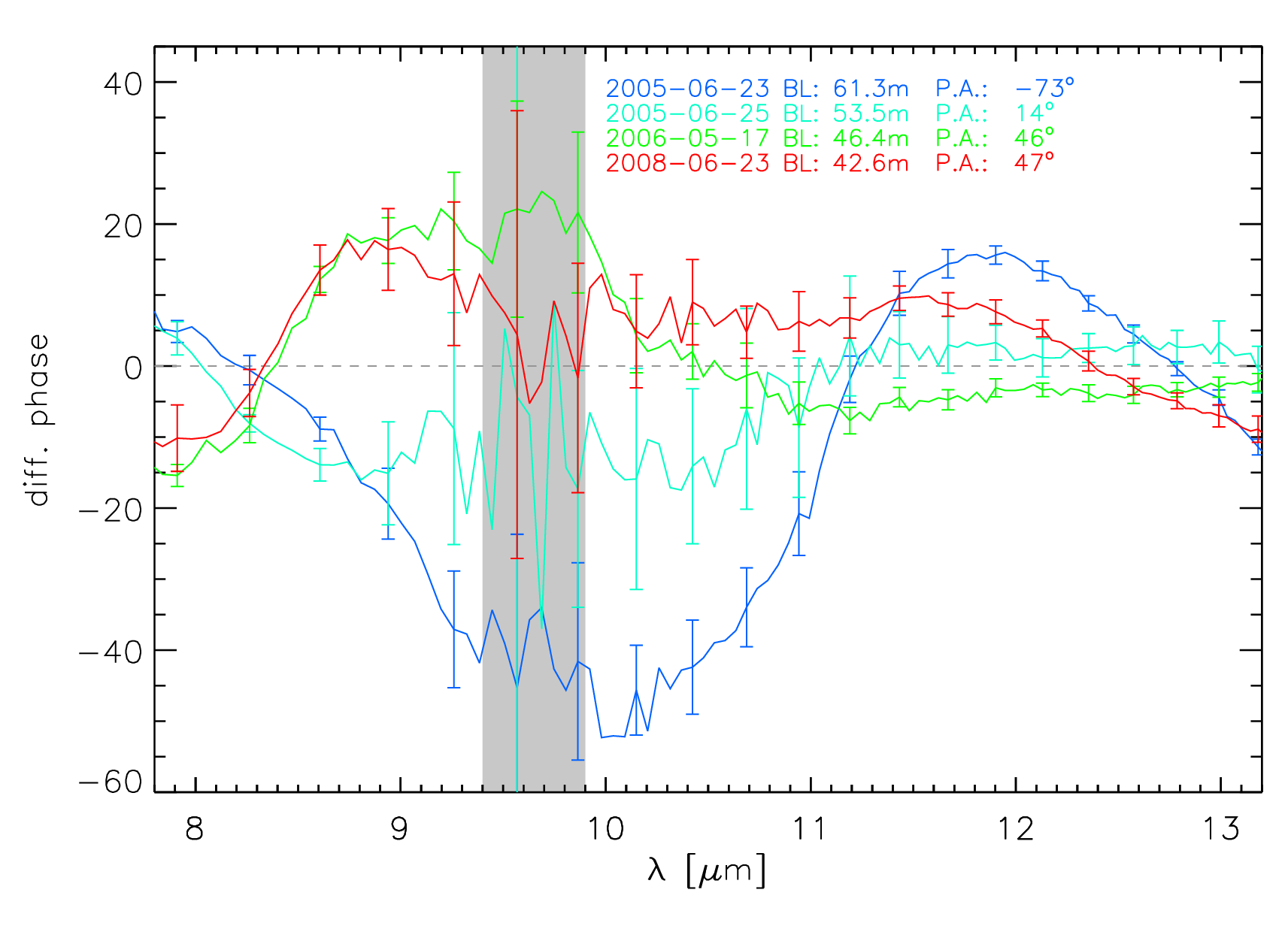}
 \caption{Direct outcome of the MIDI observations for AFGL 2136. 
        Shown are our three measurements, as well as the data 
	from \citet{2011A&A...526L...5D}, reduced by our pipeline. 
	{\bf Left}: Calibrated visibilities. {\bf Right}: Raw differential  
	phases. In both plots, the shown error bars represent    
	3$\sigma$.}\label{Fig:AFGL2136-1}
\end{figure}
The object AFGL 2136 IRS1 is the central high--mass source within the so--called Juggler Nebula \citep[e.g.,][]{1992ApJ...389..357K}, a bright near--infrared reflection nebula at an assumed distance of $\sim$2 kpc. \citet{1994ApJ...425..695K} identified a large bipolar molecular outflow emanating from this region, indirectly indicating the presence of a circumstellar disk associated with the central object. Very recently, \citet{2011A&A...526L...5D} presented a first MIDI observation of AFGL 2136 for one baseline (42.6 m), showing strongly resolved thermal emission. We have obtained three visibility points within our programme for AFGL 2136 with different UT baseline lengths and orientations. We show the visibilities and differential phases in Fig.~\ref{Fig:AFGL2136-1}.

\begin{figure}[t]
 \includegraphics[width=5.5cm]{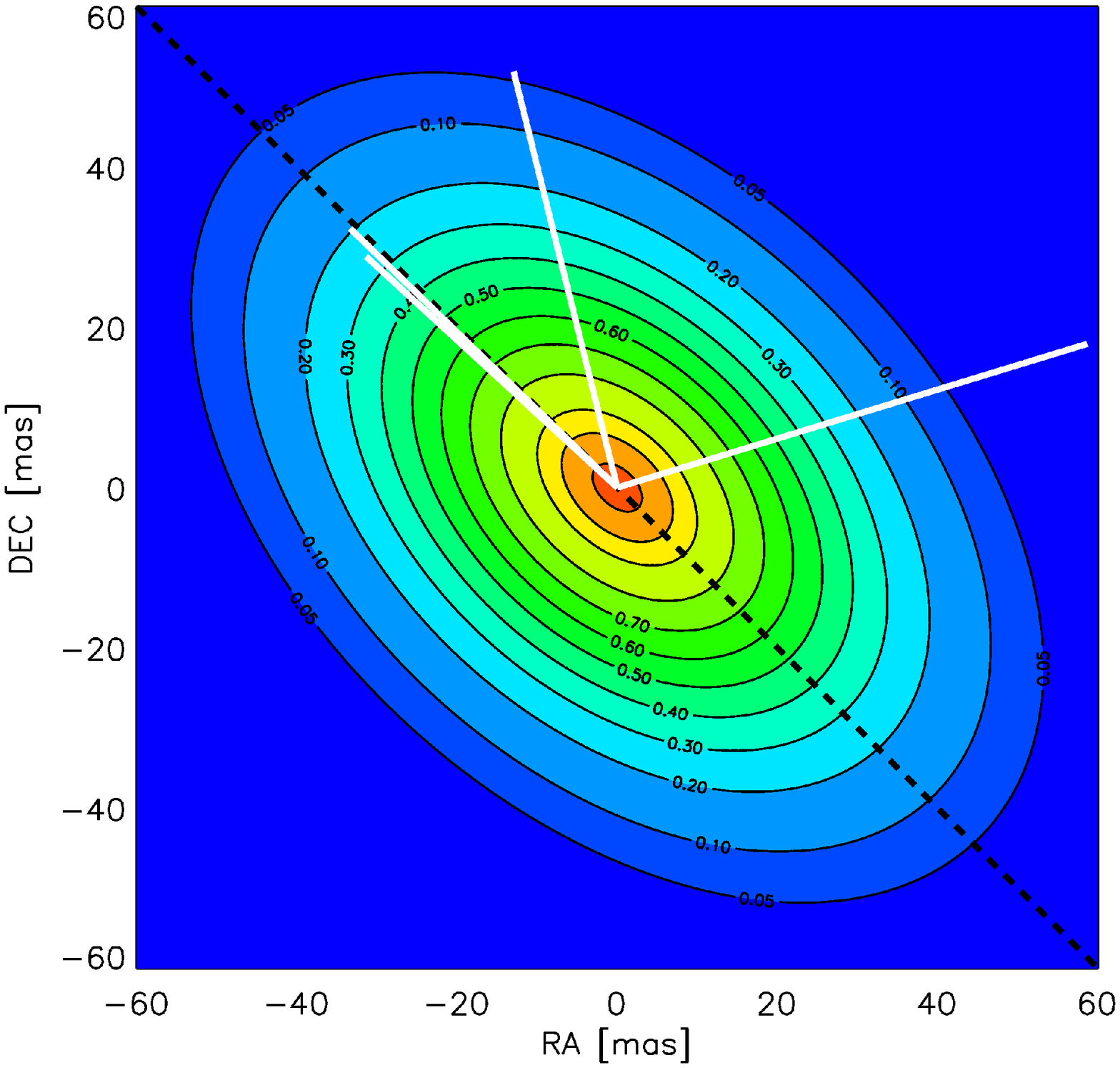}\includegraphics[width=7.5cm]{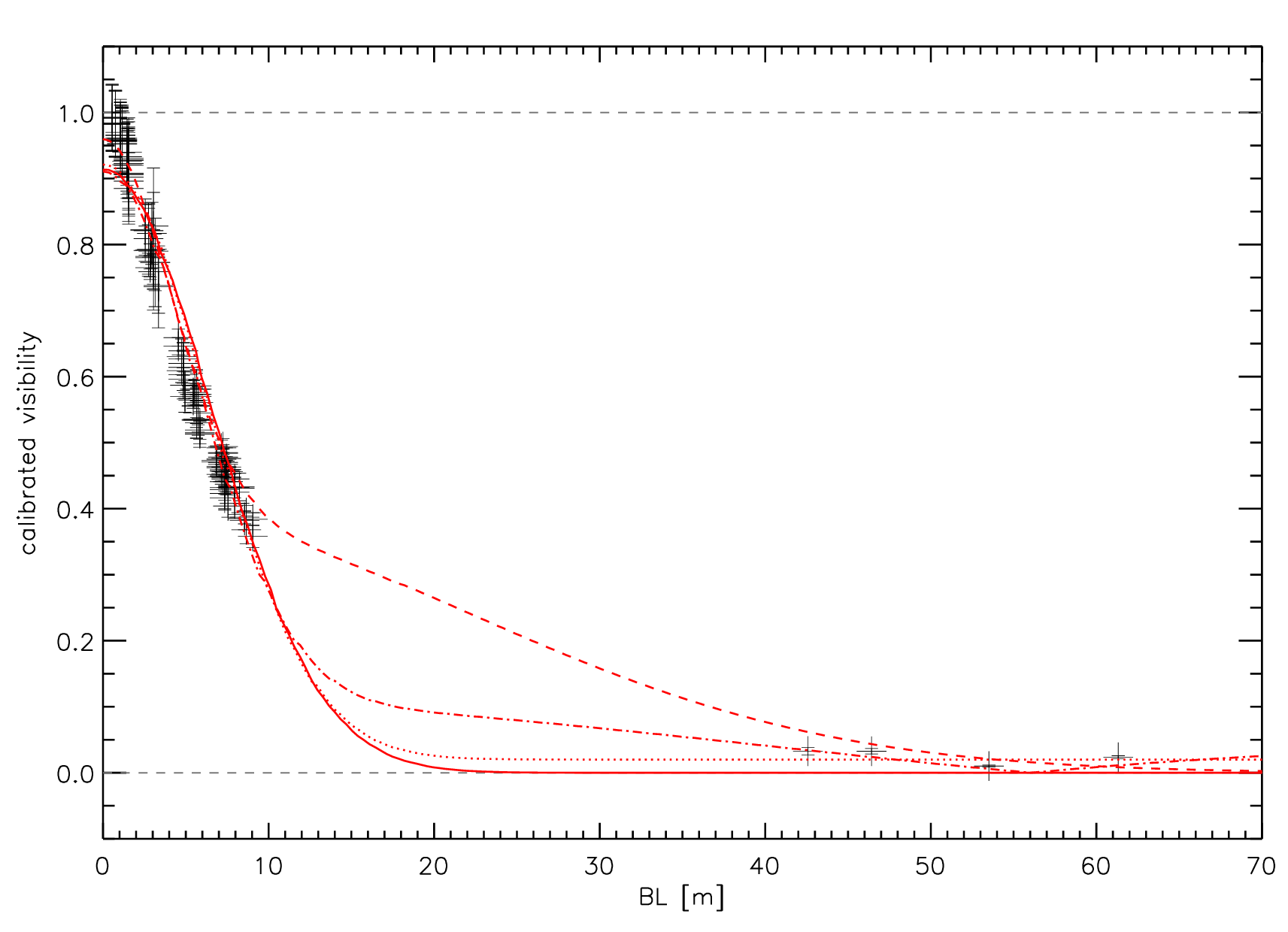}
 \caption{{\bf Left}: 2D Gaussian size fit to the measured AFGL 2136 visibilities           (whi\-te lines depicting the four MIDI baselines). {\bf Right}:         
        (u,v)--plane fits of simple analytic models. Solid: Gaussian + halo;   
	dotted: Gaussian + halo + point source; dashed: 2 Gaussians + halo; 
	dash--dotted: Gaussian + halo + ring. The black plus signs denote the 
	Keck data  (BL$\le$10 m) from \citet{2009ApJ...700..491M} and the four 
	MIDI visibilities (BL$\ge$40 m) }\label{Fig:AFGL2136-2}
\end{figure}

\noindent
Also the new MIDI data indeed show a very low visibility level, speaking for quite extended MIR emission. Although the (u,v)--coverage of the MIDI observations is sparse, it is possible to fit a 2D--Gaussian approximation to the brightness distribution probed by these data (Fig.~\ref{Fig:AFGL2136-2}, left). First, we note that the fitted position angle of the Gaussian major axis (42$\pm 6^\circ$) is almost perpendicular to the previously established bipolar outflow which is oriented from northeast to southwest of IRS 1 \citep{1994ApJ...425..695K}. Second, a comparison with the Keck segment--tilt MIR data presented by \citet{2009ApJ...700..491M} is instructive. With their baselines $\le$10 m, they were deducing an almost radially symmetric intensity distribution with a 2D--Gaussian axis ratio $<$1.1. With the 4--6 times longer MIDI baselines, we now reveal stronger deviations from radial symmetry, with axis ratios between 1.4 and 1.9 (depending on wavelength). This points to the existence of an inclined axisymmetric structure being present on scales $<$ 100 mas. These deviations from simple geometries appear to be even larger at the smallest scales we probe. The differential phases indicate, in simple terms, photo centre shifts of the brightness distribution with wavelengths. Note, that we see the largest amplitudes of these phases at the longest baseline probing linear size scales of $<$ 70--100 AU in this case. 
\noindent
In Fig.~\ref{Fig:AFGL2136-2}, right, we compare simple geometrical compound models with the Keck and the MIDI data. The Gaussian + halo model \citep[similar to the one used in][]{2009ApJ...700..491M} predicts a visibility level of basically zero for the range of the MIDI observations. Improvements can be obtained by employing two 2D Gaussians + halo, an especially by inclusion of a ring component. However, this figure demonstrates that MIDI observations in the baseline range 10\,--\,40 m would be necessary to reliably distinguish between such models. This can just be achieved by using MIDI in combination with the auxiliary telescopes (ATs). However, the scarcity of sufficiently bright optical guide stars (brightest star within 60$''$ has R=13.95 mag and B=14.88 mag) currently prevents AT observations towards AFGL 2136 (see also Sect.~\ref{Sect:MonR2}).

\noindent
The full AFGL 2136 data set, including correlated and uncorrelated fluxes as well as archival data from other facilities, is also used to investigate the properties of the dust grains resonsible for the MIR emission and absorption. We will not dwell on these subjects here, but just refer to the forthcoming paper by Follert et al.~(to be subm.) that will give much more detail on many aspects of the AFGL 2136 observations and interpretation.

\subsection{Mon R2 IRS 3}\label{Sect:MonR2}

\noindent
\begin{figure}[t]
   \hspace*{-0.35cm}\includegraphics[width=7cm]{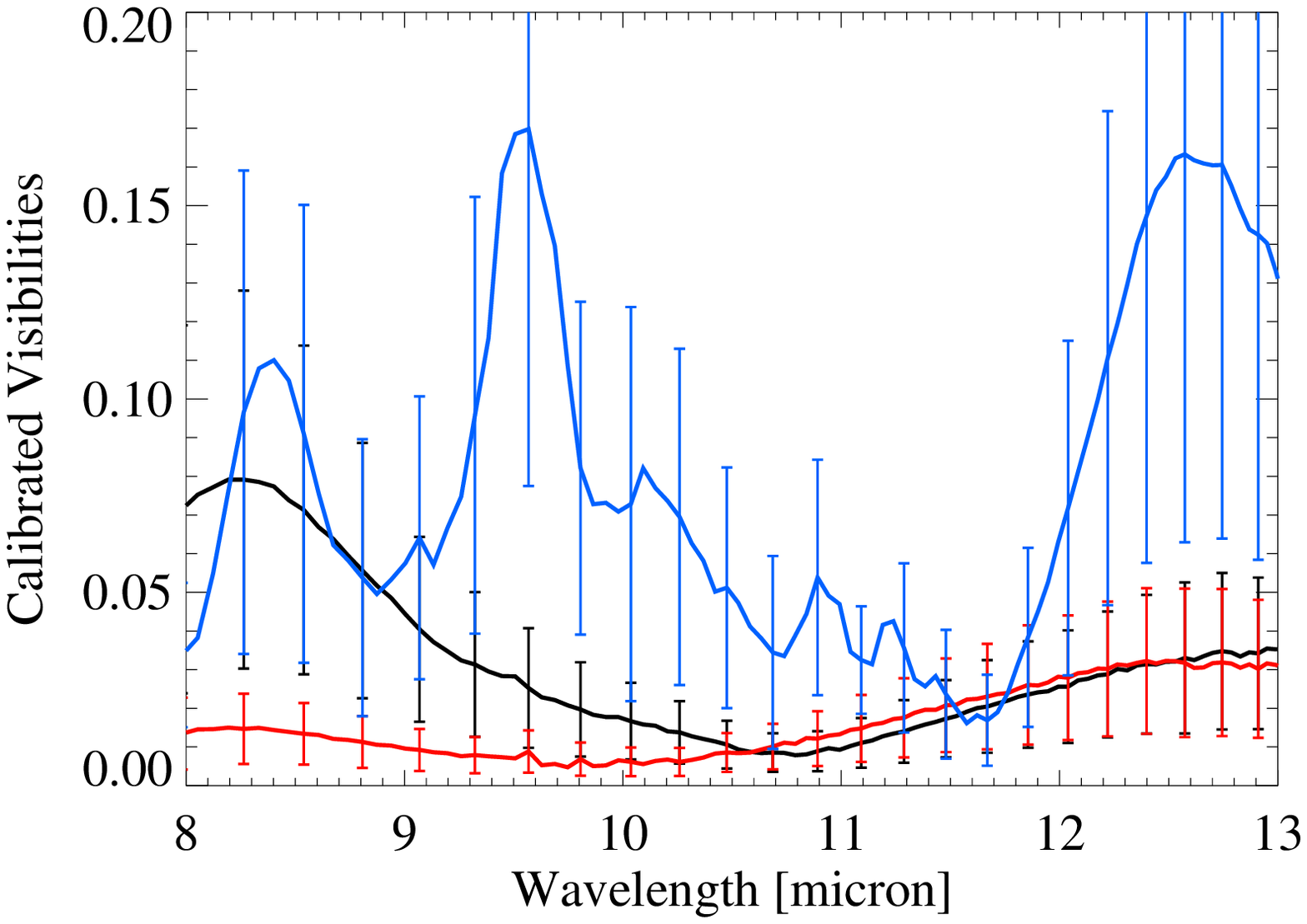}\hspace*{-0.25cm}\includegraphics[width=7cm]{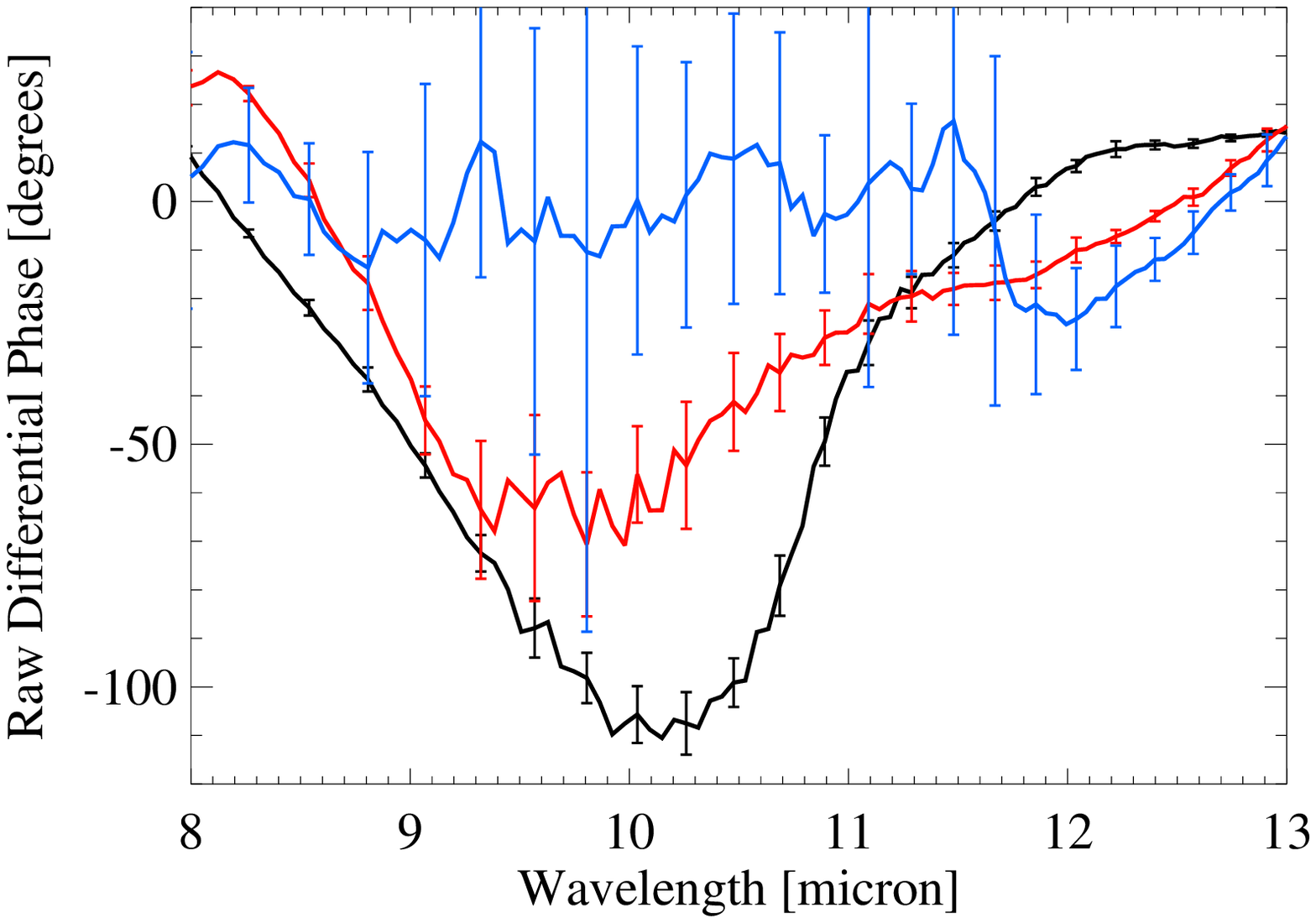}
   \caption{Direct outcome of the MIDI observations for Mon R2 IRS 3A. 
        Shown are the two UT observations in black (46.2 m, 136.3$^\circ$) and 
	red (52.9 m, 27.0$^\circ$), as well as the very noisy AT data set in 
	blue (26.6 m, 154.4$^\circ$).
	{\bf Left}: Calibrated visibilities. {\bf Right}: Raw differential  
	phases. For the visibilities, the shown error bars represent    
	1$\sigma$, for the differential phases 3$\sigma$ is 
	plotted.}\label{Fig:MonR2-1}
\end{figure}

\noindent
The Monoceros R2 region hosts several well--known massive YSOs. Among them is the IRS 3 region (at around 830 pc), which appears very bright at near-- and mid--infrared wavelengths. \citet{2002A&A...392..945P} could resolve this emission into a wealth of extended emission features and several embedded YSOs by means of NIR speckle interferometry. Dominating source is the object IRS 3A (in the Preibisch et al. nomenclature), which also appears to drive an outflow that shapes the surrounding material accordingly. This gives the NIR emission a cone--shaped appearance on scales of $<1''$.

\begin{figure}[t]
   \includegraphics[width=7.5cm]{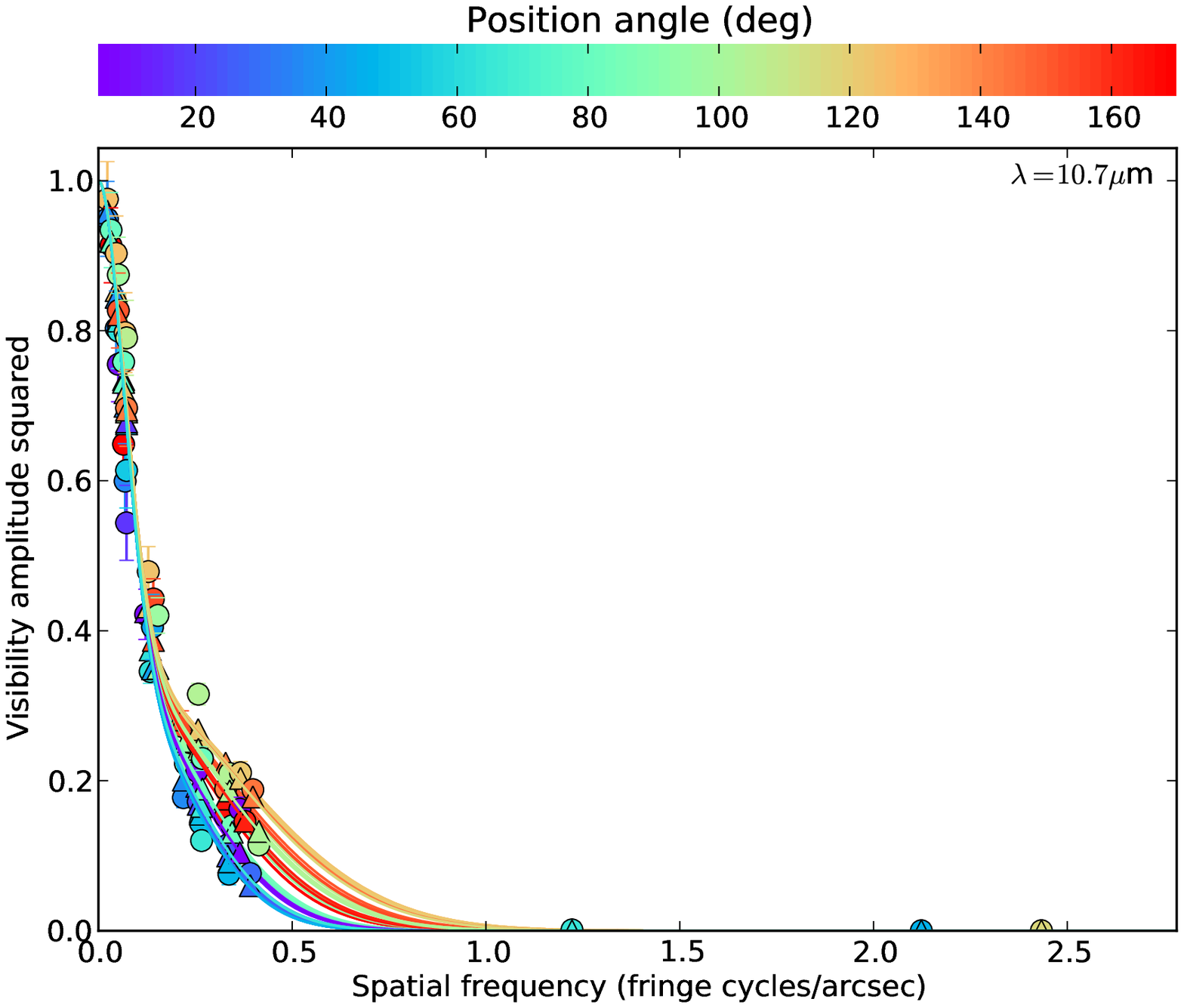}\includegraphics[width=5cm]{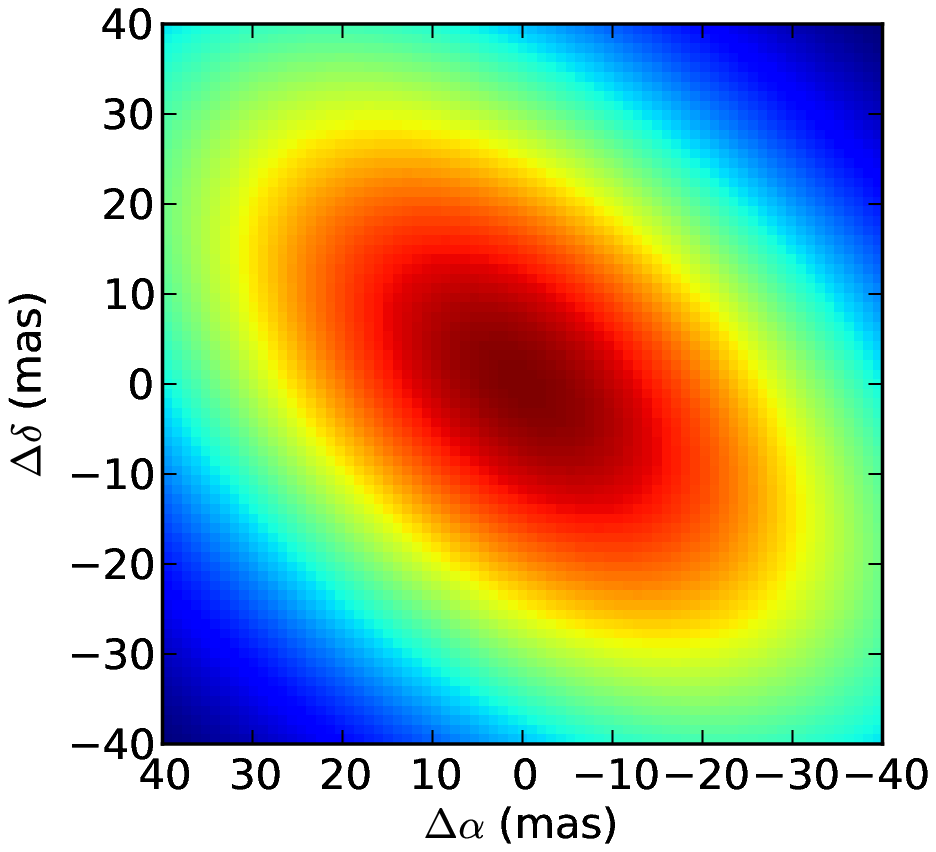}
   \caption{{\bf Left}: Squared visibility data at 10.7 $\mu$m for the Keck segment--tilt data by \citet{2009ApJ...700..491M} as well as our three MIDI observations, including simple 2D Gaussian size fits. {\bf Right}: The reconstructed intensity distribution, based on these fits.}\label{Fig:MonR2-2}
\end{figure}

\noindent
Up to now, two visibility measurements with almost perpendicular projected UT baseline orientations have been obtained. The general visibility levels, as in the case for AFGL 2136, are quite low (Fig.~\ref{Fig:MonR2-1}, left). This is expectable, since already the Keck segment--tilt data by \citet{2009ApJ...700..491M} showed resolved 10.7 $\mu$m emission for this particular source. We have performed a component size fit, using the Keck data as well as the MIDI visibilities (Fig.~\ref{Fig:MonR2-2}). We have used the combination of a 1D Gaussian (similar to a general halo component) and a 2D Gaussian. The best fit gives a relative brightness ratio of 0.59 for these two Gaussians. The 1D Gaussian is very extended (FWHM=520 mas), while the more compact 2D Gaussian has a major axis FWHM of 130 mas and an axis ratio of 1.77. This is in good agreement with the fit in \citet{2009ApJ...700..491M}, of course, since in this combined fit, the many Keck data dominate compared to the few MIDI visibilities which furthermore feature very small values only. The position angle for the 2D Gaussian in our fit is 48.8$\pm$2.5 degrees. Hence, this dominating emission component is well aligned with the outflow structure indicated by the northeast--to--southwest oriented scattering lobes in the NIR (see above). This confirms that MIR emission on sub-arcsecond scales still can be dominated by (re-)emission from the inner walls of outflow cavities, as was recently found observationally in a variety of circumstances \citep[e.g.,][]{2005A&A...429..903H,2006ApJ...642L..57D}. If a circumstellar structure exists in Mon R2 IRS 3A perpendicular to the outflow lobes, then it carries just little correlated flux on baselines probed by the Keck and our MIDI data. Nevertheless, the differential phases derived from the MIDI data show a clear and strong signal. Hence, on very compact scales they reveal additional asymmetries that do not seem to be identical with the outflow lobe structure. Indeed, the impressive differential--phase signal of $\sim 110^\circ$ occurs for the baseline roughly perpendicular to the outflow structure. Thus, a very interesting physical structure gives its imprint here, perhaps the actual circumstellar disk. 

\noindent
An attempt has been made to observe IRS 3 A with MIDI and the ATs. One useable but very noisy data set could be obtained (cf.~Fig.~\ref{Fig:MonR2-1}). However, several more attempts in service mode have failed, since the available optical guide star (V = 14.16 mag according to our own optical photometry, based on recent observations with the 2-m/1.34-m Schmidt telescope of the Karl--Schwarzschild--Observatory in Tautenburg, Germany) is actually beyond the initial guiding performance limit of the ATs (V$\le$13.5 mag). This indicates that the recently performed refurbishment of the STRAP tip--tilt units for optical guiding on the ATs gave only minor improvements in this regard. Hence, we eagerly await the employment of a new--generation adaptive--optics system on the ATs \cite[NAOMI,][]{2010SPIE.7734E...3H}. This system promises reasonable performance up to V = 15 mag which is a prerequisite for observing more than just a handful of massive YSOs with MIDI + ATs, since in many cases, the limit is currently {\it not} governed by the correlated flux levels, but by the brightness of available optical guide stars.

\section{Conclusions}

Massive YSOs can showcase a quite complicated structure, in particular, when the sources drive strong outflows. Observations with single 8--10-m telescopes might also pick up emission associated with the inner outflow cavities. MIR interferometry with MIDI traces spatial scales 4 to 10 times smaller, but it strongly depends on the individual object whether still the MIR emission from the outflow cones, the envelope, or the circumstellar disk dominates the MIDI signals. It is worth noting that circumstellar disks found around MYSOs and Herbig Be stars by millimetre interferometry tend to be small \citep[$< 100$ AU; e.g.,][]{2009A&A...497..117A}. Hence, long MIDI baselines are required to detect such disks, but low visibility levels are expected for such baselines.

\noindent
On longer MIDI baselines, we occasionally see significant differential phase signals. Hence, we eventually pick up clear deviations from spherical symmetry on scales $\la$ 40 mas. We explicitly mention that these strong and useful differential phase signals exist especially in cases where the visibilities are very low. Hence, MIDI observations scrutinising this low--visibility regime of V=0.01\,--\,0.1 can nevertheless reveal highly relevant structure information.

\acknowledgements We thank the staff of the ESO Paranal observatory and Christian Hummel at ESO Garching for their support. Thanks to John Monnier for making his Keck segment--tilting experiment data available in electronic form.

\bibliographystyle{asp2010}
\bibliography{LinzMRO}

\begin{thebibliography}{}
\expandafter\ifx\csname natexlab\endcsname\relax\def\natexlab#1{#1}\fi
\expandafter\ifx\csname url\endcsname\relax
  \def\url#1{\texttt{#1}}\fi
\expandafter\ifx\csname urlprefix\endcsname\relax\def\urlprefix{URL }\fi
\providecommand{\eprint}[2][]{\url{#2}}

\bibitem[{{Alonso-Albi} et~al.(2009){Alonso-Albi}, {Fuente}, {Bachiller},
  {Neri}, {Planesas}, {Testi}, {Bern{\'e}}, \& {Joblin}}]{2009A&A...497..117A}
{Alonso-Albi}, T., {Fuente}, A., {Bachiller}, R., {Neri}, R., {Planesas}, P.,
  {Testi}, L., {Bern{\'e}}, O., \& {Joblin}, C. 2009, \aap, 497, 117.
  \eprint{0812.1636}

\bibitem[{{Beuther} et~al.(2007){Beuther}, {Churchwell}, {McKee}, \&
  {Tan}}]{2007prpl.conf..165B}
{Beuther}, H., {Churchwell}, E.~B., {McKee}, C.~F., \& {Tan}, J.~C. 2007,
  Protostars and Planets V, 165. \eprint{arXiv:astro-ph/0602012}

\bibitem[{{De Buizer}(2006)}]{2006ApJ...642L..57D}
{De Buizer}, J.~M. 2006, \apjl, 642, L57. \eprint{arXiv:astro-ph/0603428}

\bibitem[{{de Wit} et~al.(2011){de Wit}, {Hoare}, {Oudmaijer},
  {N{\"u}rnberger}, {Wheelwright}, \& {Lumsden}}]{2011A&A...526L...5D}
{de Wit}, W.~J., {Hoare}, M.~G., {Oudmaijer}, R.~D., {N{\"u}rnberger},
  D.~E.~A., {Wheelwright}, H.~E., \& {Lumsden}, S.~L. 2011, \aap, 526, L5+.
  \eprint{1012.0179}

\bibitem[{{Follert} et~al.(2010){Follert}, {Linz}, {Stecklum}, {van Boekel},
  {Henning}, {Feldt}, {Herbst}, \& {Leinert}}]{2010A&A...522A..17F}
{Follert}, R., {Linz}, H., {Stecklum}, B., {van Boekel}, R., {Henning}, T.,
  {Feldt}, M., {Herbst}, T.~M., \& {Leinert}, C. 2010, \aap, 522, A17+.
  \eprint{1007.4079}

\bibitem[{{Haguenauer} et~al.(2010){Haguenauer}, {Alonso}, {Bourget}, \& {et
  al.}}]{2010SPIE.7734E...3H}
{Haguenauer}, P., {Alonso}, J., {Bourget}, P., \& {et al.} 2010, in Optical and
  Infrared Interferometry II. Proc.~SPIE, Vol.~7734, edited by {W.~C.~Danchi,
  F.~Delplancke, \& J.~K.~Rajagopal}, 773404

\bibitem[{{Henning}(1990)}]{1990FCPh...14..321H}
{Henning}, T. 1990, Fundamentals of Cosmic Physics, 14, 321

\bibitem[{{Kastner} et~al.(1992){Kastner}, {Weintraub}, \&
  {Aspin}}]{1992ApJ...389..357K}
{Kastner}, J.~H., {Weintraub}, D.~A., \& {Aspin}, C. 1992, \apj, 389, 357

\bibitem[{{Kastner} et~al.(1994){Kastner}, {Weintraub}, {Snell}, {Sandell},
  {Aspin}, {Hughes}, \& {Baas}}]{1994ApJ...425..695K}
{Kastner}, J.~H., {Weintraub}, D.~A., {Snell}, R.~L., {Sandell}, G., {Aspin},
  C., {Hughes}, D.~H., \& {Baas}, F. 1994, \apj, 425, 695

\bibitem[{{Leinert} et~al.(2003){Leinert}, {Graser}, {Waters}, \& {et
  al.}}]{2003SPIE.4838..893L}
{Leinert}, C., {Graser}, U., {Waters}, L.~B.~F.~M., \& {et al.} 2003, in
  Interferometry for Optical Astronomy II. Proc.~SPIE, Vol.~4838, edited by
  W.~A. {Traub}, 893

\bibitem[{{Linz} et~al.(2009){Linz}, {Henning}, {Feldt}, {Pascucci}, {van
  Boekel}, {Men'shchikov}, {Stecklum}, {Chesneau}, {Ratzka}, {Quanz},
  {Leinert}, {Waters}, \& {Zinnecker}}]{2009A&A...505..655L}
{Linz}, H., {Henning}, T., {Feldt}, M., {Pascucci}, I., {van Boekel}, R.,
  {Men'shchikov}, A., {Stecklum}, B., {Chesneau}, O., {Ratzka}, T., {Quanz},
  S.~P., {Leinert}, C., {Waters}, L.~B.~F.~M., \& {Zinnecker}, H. 2009, \aap,
  505, 655. \eprint{0907.0445}

\bibitem[{{Linz} et~al.(2008){Linz}, {Stecklum}, {Follert}, {Henning}, {van
  Boekel}, {Men'shchikov}, {Pascucci}, \& {Feldt}}]{2008JPhCS.131a2024L}
{Linz}, H., {Stecklum}, B., {Follert}, R., {Henning}, T., {van Boekel}, R.,
  {Men'shchikov}, A., {Pascucci}, I., \& {Feldt}, M. 2008, Journal of Physics
  Conference Series, 131, 012024. \eprint{0809.1384}

\bibitem[{{Linz} et~al.(2005){Linz}, {Stecklum}, {Henning}, {Hofner}, \&
  {Brandl}}]{2005A&A...429..903H}
{Linz}, H., {Stecklum}, B., {Henning}, T., {Hofner}, P., \& {Brandl}, B. 2005,
  \aap, 429, 903. \eprint{arXiv:astro-ph/0406680}

\bibitem[{{Monnier} et~al.(2009){Monnier}, {Tuthill}, {Ireland}, {Cohen},
  {Tannirkulam}, \& {Perrin}}]{2009ApJ...700..491M}
{Monnier}, J.~D., {Tuthill}, P.~G., {Ireland}, M., {Cohen}, R., {Tannirkulam},
  A., \& {Perrin}, M.~D. 2009, \apj, 700, 491. \eprint{0905.3495}

\bibitem[{{Preibisch} et~al.(2002){Preibisch}, {Balega}, {Schertl}, \&
  {Weigelt}}]{2002A&A...392..945P}
{Preibisch}, T., {Balega}, Y.~Y., {Schertl}, D., \& {Weigelt}, G. 2002, \aap,
  392, 945

\end{thebibliography}

\end{document}